%% file: main.tex
\documentclass[]{article}
\usepackage{amsmath,amsfonts}
\usepackage{textcomp}
\usepackage{url}
\usepackage{graphicx}

\usepackage{multirow}
\usepackage{color}
\usepackage[margin=20mm]{geometry}

\begin{document}
\title{\textbf{RISC: A Corpus for Shout Type Classification\\ and Shout Intensity Prediction}} 
\date{}
\author{Takahiro Fukumori$^{\dag}$, Taito Ishida$^{\dag\dag}$, and Yoichi Yamashita$^{\dag}$\\
{\normalsize$^{\dag}$College of Information Science and Engineering, Ritsumeikan University, Japan.}\\
{\normalsize$^{\dag\dag}$Graduate School of Information Science and Engineering, Ritsumeikan University, Japan.}
}
\maketitle

\begin{abstract}
The detection of shouted speech is crucial in audio surveillance and monitoring.
Although it is desirable for a security system to be able to identify emergencies, existing corpora provide only a binary label (i.e., shouted or normal) for each speech sample, making it difficult to predict the shout intensity. 
Furthermore, most corpora comprise only utterances typical of hazardous situations, meaning that classifiers cannot learn to discriminate such utterances from shouts typical of less hazardous situations, such as cheers. 
Thus, this paper presents a novel research source, the RItsumeikan Shout Corpus (RISC), which contains wide variety types of shouted speech samples collected in recording experiments. 
Each shouted speech sample in RISC has a shout type and is also assigned shout intensity ratings via a crowdsourcing service.
We also present a comprehensive performance comparison among deep learning approaches for speech type classification tasks and a shout intensity prediction task.  
The results show that feature learning based on the spectral and cepstral domains achieves high performance, no matter which network architecture is used.
The results also demonstrate that shout type classification and intensity prediction are still challenging tasks, and RISC is expected to contribute to further development in this research area. 
\end{abstract}

\input{sec1-introduction}
\input{sec2-related}
\input{sec3-dataset}
\input{sec4-proposed}
\input{sec5-evaluation}
\input{sec6-conclusion}

\section*{Acknowledgments}
This work was supported by JSPS KAKENHI Grant Number JP21K14381.
This study was approved by the research ethics committee of Ritsumeikan University (permission number: BKC-LSMH-2021-081).

\bibliographystyle{IEEEtran}
\bibliography{main}
\vfill
\end{document}

%% file: sec1-introduction.tex
\section{Introduction}
\label{sec:intro}
The development of automated surveillance systems is essential to protect people's safety.
To date, computer vision techniques have often been applied to video data captured by cameras~\cite{19-wang-tifs,19-singh-tits}.
Recently, many studies have also focused on the use of audio information recorded by microphones for abnormal situation detection~\cite{16-crocco-acm}.
Typical examples of sound categories targeted by conventional research include gunshots~\cite{21-rahman-mta}, alarms~\cite{17-carmel-eusipco}, rainfall~\cite{22-trucco-joe}, running vehicles~\cite{18-almaadeed-sensors,18-li-access}, and mechanical faults~\cite{21-wichern-waspaa}.
In addition to the detection of such audio events, the ability to distinguish shouted speech from ordinary speech such as daily conversations is highly useful for emergency rescue operations. 
This problem can be formulated as a specific type of speech classification in which an input speech sample is judged as either a shout or not. 
In several studies, a labeled corpus comprising shouted and normal speech has been constructed as a training resource~\cite{17-eusipco-valenti,20-gaviria-as,21-fukumori-interspeech,10-huang-ciea,10-chan-eurosip}.

As the basis for a practical system of audio surveillance, the conventional corpora used in the literature are insufficient for two reasons.
First, the existing corpora contain only binary labels for speech samples, i.e., shouted or normal, rather than a numerical score indicating the shout intensity. 
Although an audio surveillance system should ideally be able to judge different instances to assign different priorities for rescue, it is not straightforward to compare the level of emergency between shouts based on learning from binary labels only.
To solve this problem, each shouted speech sample should be associated with a \textit{shout intensity}---the degree of `shout-like-ness' perceived by the listener. 
Here, we should emphasize that the shout intensity cannot be quantified using the sound pressure level of the speech because the sound pressure level greatly depends on the positional relationship between the microphone and the speaker.
Second, most of the existing corpora comprise only utterances that typically occur in emergency situations (e.g., ``help!''~\cite{20-gaviria-as}). 
However, people also often shout for joy in nonhazardous situations. 
Although an audio surveillance system must discriminate between these different shout types, conventional studies have ignored this fact, and the feasibility of such discrimination is still unknown. 
The ability to predict a speaker's situation (i.e., hazardous or not) and emergency level will require a new shouted speech corpus labeled with shout type and intensity information.

This paper presents a novel corpus of shouted speech, the \textbf{RI}tsumeikan \textbf{S}hout \textbf{C}orpus (\textbf{RISC}), comprising angry shouts, screams, and cheers collected from a recording experiment at Ritsumeikan University. 
The process of creating of our corpus started with defining a list of possible sentences of shouted speech.
Then, we asked experiment participants to utter each sentence while imagining a situation for which the sentence would be suitable.
Finally, based on listening experiments using a crowdsourcing service, each shouted speech sample was assigned shout intensity ratings as crucial information for training emergency detectors.

This paper also considers how to predict the speech type or shout intensity for a given speech sample.
In recent years, deep learning has become a mainstream approach to shouted speech detection. 
For example, several methods have been proposed to model the relationship between the temporal variations of speech features and the speech status using convolutional neural networks (CNNs) or recurrent neural networks (RNNs)~\cite{17-eusipco-valenti,20-gaviria-as,21-fukumori-interspeech}. 
Most conventional studies have used traditional, manually designed low-dimensional features as the input to these networks. 
Typical features of this type include the mel-frequency cepstral coefficients (MFCCs)~\cite{20-gaviria-as} and the mel spectrogram~\cite{20-jasa-baghel}.
Here, we focus on the fact that other recent speech processing tasks have shown the effectiveness of automatic feature extraction from high-dimensional information. 
For example, temporal waveforms~\cite{20-parcollet-icassp} and spectrograms~\cite{21-fan-taslp} have been shown to improve the speech recognition performance of deep learning compared with traditional low-dimensional features. 
This trend can also be seen for speaker identification, for which deep models can automatically extract effective features from input raw speech~\cite{18-jung-icassp}.
Following these works, our recent study~\cite{21-fukumori-interspeech} has presented a novel speech classification method based on spectrogram and cepstrogram features obtained by arranging the spectra and cepstra, respectively, as time series.
In this paper, we present a comprehensive performance comparison between the conventional methods and our deep spectral--cepstral approach~\cite{21-fukumori-interspeech} based on RISC for not only classification but also regression.

The main contributions of this paper are summarized as follows:

\begin{itemize}
    \item We have constructed a novel corpus with various shout types and shout intensity ratings, which can support new recognition challenges in shouted speech detection research. This paper describes the details of its construction pipeline, including in-laboratory speech recording and crowdsourcing-based verification. 
    Furthermore, we have released our corpus on the web\footnote{\url{https://t-fukumori.net/corpus/RISC/en.html}}.
    \item Using the constructed corpus, we present comprehensive results for classification and regression obtained with conventional methods and our deep spectral--cepstral approach.
\end{itemize}

The remainder of this paper is organized as follows. 
Section~\ref{sec:related} introduces conventional deep approaches for shout detection and the existing speech corpora used for model training.
Section~\ref{sec:dataset} describes the procedure used to construct our corpus. 
In particular, we explain how we recorded the shouted speech samples and obtained the intensity ratings for each sample.
Section~\ref{sec:proposed} describes the acoustic features and the structures of the deep approaches for detecting shouted speech.
In Section~\ref{sec:evaluation}, we present the results of experiments on shouted vs. normal speech classification, shout type classification, and shout intensity prediction based on RISC.
Finally, Section~\ref{sec:conclusion} concludes the paper and suggests some possible directions for future work.

%% file: sec2-related.tex
\section{Related works}
\label{sec:related}
\subsection{Deep approaches for shouted speech detection}
Deep neural networks (DNNs) have dramatically improved the performance of speech analysis technology in recent years.
Additionally, for shouted speech detection, DNNs have been shown to outperform conventional classifiers, such as Gaussian mixture models and hidden Markov models~\cite{11-pohjalainen-icassp, 09-ntalampiras-eurasip, 09-nanjo-isce,09-ntalampiras-icassp}.
Therefore, a recent focus of related research has been the design of features that are suitable as inputs to DNNs.
For example, Laffitte et al.~\cite{16-laffitte-icassp} used the MFCCs and energy components to train a deep architecture consisting of restricted Boltzmann machines and deep belief networks for shouted speech detection.
Baghel et al.~\cite{20-jasa-baghel} also calculated the MFCCs and their second derivatives for use as inputs to a DNN.
Gaviria et al.~\cite{20-gaviria-as} used the MFCCs and the mel spectrogram in their deep learning model.
A recent method presented by Baghel et al.~\cite{21-baghel-interspeech} relied on calculating an integrated linear prediction residual~\cite{13-prathosh-taspl}, representing the period information of vocal fold vibration.
The network architecture used in~\cite{21-baghel-interspeech} consisted of an autoencoder, an attention mechanism, and bidirectional gated recurrent units (GRUs).

Shouted speech is sometimes considered as one of the target classification categories for environmental sound recognition~\cite{21-mesaros-ieee,17-icot-dang}.
For example, Mun et al.~\cite{17-mun-icassp} used the MFCCs to train DNNs in experiments on a home surveillance environment database.
Valenti et al.~\cite{17-eusipco-valenti} also detected acoustic sound events using DNNs and RNNs; their features were the log-mel spectrogram and its first derivative.
These previous studies also used traditional and handcraft speech features. 

Our recent work~\cite{21-fukumori-interspeech} presented a novel approach based on learning descriptive features from the spectral and cepstral domains for shouted speech detection.
Specifically, we used two types of high-dimensional features, spectrograms and cepstrograms, as inputs to a deep architecture.
This feature learning approach showed superior performance over conventional low-dimensional features.
As the major difference between the present paper and the previous conference version~\cite{21-fukumori-interspeech},
this paper presents a comprehensive performance comparison among different deep approaches on our new corpus.
Furthermore, whereas previous studies, including our recent work~\cite{21-fukumori-interspeech}, have addressed only the classification task, this paper reports the performance for not only classification but also shout intensity prediction. 

\subsection{Existing corpora for shouted speech detection}
A training corpus is essential for developing a shouted-speech detector, and many research groups have constructed their own corpora for this purpose. 
Table~\ref{tbl:corpus-list} presents comparisons between existing corpora and our new corpus.
Nandwana et al.~\cite{14-nandwana-interspeech} recorded screams and neutral speech samples for binary classification.
The corpus used in~\cite{16-laffitte-icassp} contained shouted speech collected in subway trains, with each speech sample labeled as ``scream,'' ``shout,'' ``conversation,'' or ``noise.''
Mesbahi et al.~\cite{19-mesbahi-ijst} collected 91 shouted speech samples from various web sources, including screams, expressions of panic, baby cries, cries of pain, etc., for speech characteristic analysis. 
The corpus constructed in~\cite{20-jasa-baghel} comprised Indian English utterances with binary labels (i.e., normal and shouted).
In~\cite{21-baghel-interspeech}, the authors subsequently presented a speech dataset taken from news debates in Indian English, with categories of normal speech, shouted speech, and noise.
Notably, in these works, the situation in which a person is shouting is usually assumed to be unknown or a hazardous one only.
An exception is a Finnish corpus presented by Pohjalainen et al.~\cite{13-pohjalainen-jasa}, in which the speakers uttered and shouted general sentences that could be encountered in both hazardous and nonhazardous. 
However, they used no situation labels for classifier training, and there was no discussion of shout type classification.

Some corpora targeting environmental sound or emotional speech recognition also contain shouted speech as one of the classification categories. 
For example, an environmental sound corpus~\cite{16-mesaros-eusipco} that was used in the DCASE 2016 Challenge Task 3~\cite{16-decase2016} includes samples of children's shouts in a residential area. 
P. Foggia et al.~\cite{15-foggia-prl} collected four kinds of audio clips, namely, screams, glass breaking, gunshots, and background noise, as the main categories for audio surveillance.
Hsu et al.~\cite{21-hsu-taspl} assigned labels of either ``laughter,'' ``breathing,'' ``shout,'' or ``background'' to a subset of the speech samples in the emotional speech corpus named NNIME~\cite{17-chou-acii}.

The speech samples in the above corpora have categorical labels but no continuous values; consequently, they cannot support regressor training.
Furthermore, the corpus developers sometimes intentionally screened speech samples during their recording experiments, although their guidelines or reasons are mostly unknown.
For example, raised voices were not admitted as shouting in the corpus construction process of~\cite{20-jasa-baghel}. 
In~\cite{13-pohjalainen-jasa}, when the sound pressure level difference between a speaker's normal and shouted speech instances did not exceed a predefined threshold, the speaker's utterances were rerecorded.  
A recording engineer also instructed the speaker to repeat the utterance until they clearly recognized that the speaker was shouting. 
In our study, we avoided the influence of any single person's subjectivity by asking crowdsourcing workers to rate the intensity of each shouted speech sample, and we determined the shout types based on the speaker's intentions. 
Furthermore, where most conventional corpora remain undisclosed, we have made our corpus available on the web.

\begin{table}[ht]
  \caption{Comparison of existing corpora and our new corpus. ``Unknown'' indicates that the information is not clearly described in the literature. ``Hazardous'' indicates a hazardous situation, and ``Mixed'' comprises both hazardous and less-hazardous situations.}
  \vspace{5pt}
  \label{tbl:corpus-list}
  \centering
    \footnotesize
  \begin{tabular}{l|c|c|c|c|c|c}\hline\hline
    & \multirow{2}{*}{Language} & \multirow{2}{*}{Labels} & \multirow{2}{*}{Speech source} & Shout & Shout & Publicly\\
    & & & & situation & intensity & available\\\hline \hline
    Baghel et al.~\cite{20-jasa-baghel} & English & Normal, shout & Original recording & Unknown & & \checkmark \\\hline
    Laffitte et al.~\cite{16-laffitte-icassp} & Unknown & Scream, shout, conversation, noise & Original recording & Hazardous & & \\\hline
    Baghel et al.~\cite{21-baghel-interspeech} & English & Normal, shout, miscellaneous & News debates & Unknown & & \\\hline
    Mittal et al.~\cite{13-mittal-jasa} & English & Normal, soft, whisper, loud, shout & Original recording & Hazardous & & \checkmark \\\hline
    Pohjalainen et al.~\cite{13-pohjalainen-jasa} & Finnish & Normal, shout & Original recording & Mixed & & \\\hline 
    Mesbahi et al.~\cite{19-mesbahi-ijst} & English & Speech, shout & Several web sources & Hazardous & & \\\hline  
    \multirow{2}{*}{Mesaros et al.~\cite{16-mesaros-eusipco}} & \multirow{2}{*}{Unknown} & 18 acoustic scene classes & \multirow{2}{*}{Original recording} & \multirow{2}{*}{Unknown} & & \multirow{2}{*}{\checkmark} \\
    & & including children shouting & & & & \\\hline 
    \multirow{2}{*}{Hsu et al.~\cite{21-hsu-taspl}} & \multirow{2}{*}{Chinese} & Laughter, breathing, & \multirow{2}{*}{NNIME~\cite{17-chou-acii}} & \multirow{2}{*}{Unknown} & & \multirow{2}{*}{} \\
    & & shout, background & & & & \\\hline 
    Nandwana et al.~\cite{14-nandwana-interspeech} & English & Neutral speech, scream & Original recording & Unknown & & \\\hline
    \multirow{2}{*}{Foggia et al.~\cite{15-foggia-prl}} & \multirow{2}{*}{Unknown} & Glass breaking, gunshot, & \multirow{2}{*}{Original recording} & \multirow{2}{*}{Unknown} & & \multirow{2}{*}{\checkmark} \\
    & & scream, background noise & & & & \\\hline
    \multirow{3}{*}{\textbf{RISC (Ours)}} & \multirow{3}{*}{Japanese} & Normal, hazardous shout, & \multirow{3}{*}{Original recording} & \multirow{3}{*}{Mixed} & \multirow{3}{*}{\checkmark} & \multirow{3}{*}{\checkmark} \\
    & & less hazardous shout, & & & & \\
    & & difficult-to-classify shout & & & & \\\hline\hline
\end{tabular}
\end{table}

%% file: sec3-dataset.tex
\section{RItsumeikan Shout Corpus}
\label{sec:dataset}

This section describes the details of the creation of our corpus. 
We first listed a set of sentences to be uttered (see Section~\ref{ssec:dataset:list}).
Then, speakers uttered each sentence under our instruction (see Section~\ref{ssec:dataset:record}).
Finally, each utterance was evaluated for the shout intensity via crowdsourcing (see Section~\ref{ssec:dataset:score}).

\subsection{Listing sentences to be uttered}
\label{ssec:dataset:list}
Corpus developers usually provide speakers with ``scripts'' to smoothly conduct recording experiments.
Similarly, we prepared a set of sentences to be uttered as follows.
We first asked a group of five graduate students (four male and one female) who were engaged in spoken language research to list 55 possible sentences that people could shout in general. 
These candidate sentences were then presented to five male undergraduate students who did not have specialized knowledge of spoken language. 
They were asked whether each sentence is appropriate for shouting, and most students disagreed on only two sentences. 
We eliminated these two sentences from the candidate list, leaving 53 sentences.

Shouts can occur in various situations, and it is desirable for surveillance systems to be able to discriminate between screams and cheers.
Therefore, we asked the five undergraduate students to evaluate whether each sentence among the candidates could be uttered in a hazardous situation.
Specifically, for each of the 53 sentences, they were asked to select one impression from among three classes: ``sentence specific to highly hazardous situations (hereinafter, the \textbf{H class}),'' ``sentence specific to less hazardous situations (hereinafter, the \textbf{L class}),'' and ``sentence that is difficult to classify into hazardous or less hazardous (hereinafter, the \textbf{H/L class}).''
Based on these responses, we labeled each sentence with the class that received the largest number of votes among the three classes. 
If two classes tied for first place, the sentence was classified as belonging to the H/L class.
For each class, the 53 sentences were sorted in descending order based on the number of votes, and only the top-ranked sentences were extracted; specifically, we selected 20 sentences for the H class (e.g., ``help,'' ``shut up,'' etc., in Japanese), 20 sentences for the L class  (e.g., ``go,'' ``yes,'' etc., in Japanese), and five sentences for the H/L class (e.g., ``really?,'' ``hey,'' etc., in Japanese). 
Finally, five vowels were added to the H/L-class sentences, resulting in a list of 50 sentences for utterance. 
The RISC webpage lists all the sentences.

\subsection{Speech recording}
\label{ssec:dataset:record}
We recruited 50 graduate and undergraduate students (29 male and 21 female) and asked them to utter the 50 sentences in two different utterance styles: normal and shouted. 
After explaining the purpose and use of the recordings to the participants, we obtained informed consent from each participant. 
All speech samples were recorded in a studio with the characteristics described in Table~\ref{tbl:recording_condition}-A. 
The speaker position was 0.5 m away from the microphone installed at the center of the studio, and the microphone's vertical position was the same as that of the speaker's mouth. 
Table~\ref{tbl:recording_condition}-B lists the recording equipment and conditions. 
We fixed the input level during recording so as to prevent clipping even at 110 dBA.
Before the main recording, each participant conducted a 10-minute test recording to practice their utterances.

\begin{table}[ht]
    \caption{Conditions for speech recording.}
    \vspace{5pt}
    \label{tbl:recording_condition}
    \centering
    \begin{tabular}{l|l}\hline\hline
        \multicolumn{2}{l}{\textbf{A. Properties of the recording studio}}\\\hline
        Room size & Width: 3.1~m, Depth: 5.4~m, Height: 2.7~m\\\hline
        Reverberation time & $T_{60}$ = 200~ms\\\hline
        Temperature & \( 22~ ^\circ \)C\\\hline
        Humidity & 35\%\\\hline
        Ambient noise level & 30.0~dBA\\\hline\hline
        \multicolumn{2}{l}{\textbf{B. Recording equipment and conditions}}\\\hline
        Microphone & AKG C214\\\hline
        Audio interface & Roland Rubix22\\\hline
        Software & Audacity (ver. 2.4.2)\\\hline
        Sampling frequency & 48~kHz\\\hline
        Quantization & 16~bit\\\hline
        File format & WAV~(little endian)\\\hline\hline
    \end{tabular}
\end{table}

During the main recording, each participant first uttered the 50 sentences as normal speech and then shouted the same sentences. 
We provided a 3-minute break, including rehydration, every 25 sentences during the normal speech recordings.
A similar intermission was given every ten sentences during the shouted speech recordings to allow the participants to rest their throats.
We instructed the speakers to imagine the situations in which they were shouting when shouting the sentences in the H and L classes. 
However, we gave no special instructions for the H/L-class sentences; the speakers were allowed to shout freely. 
We also provided no example or objective criteria for including emotion in the utterances, and the speakers were allowed to act out shouting situations that they considered suitable for a given type of utterance in the H, L, and H/L classes.
During the recording experiments, speech was rerecorded only when the speaker wished it, the speaker misspoke a sentence, or a recording accident occurred. 
A total of 5,000 speech samples were collected, including 2,500 instances each of normal and shouted speech.
Furthermore, we divided all speech samples in RISC into the following four classes: (i) \textbf{Normal}: 2,500 normal speech samples, consisting of utterances of all 50 sentences; (ii) \textbf{Shout-H}: 1,000 shouted speech samples, consisting of utterances of the 20 sentences in the H class; (iii) \textbf{Shout-L}: 1,000 shouted speech samples, consisting of utterances of the 20 sentences in the L class; and (iv) \textbf{Shout-H/L}: 500 shouted speech samples, consisting of utterances of the five vowels and five sentences in the H/L class. 
These can be used for shout type classification.

\begin{figure}[ht]
    \centering
    \includegraphics[width=1.0\columnwidth]{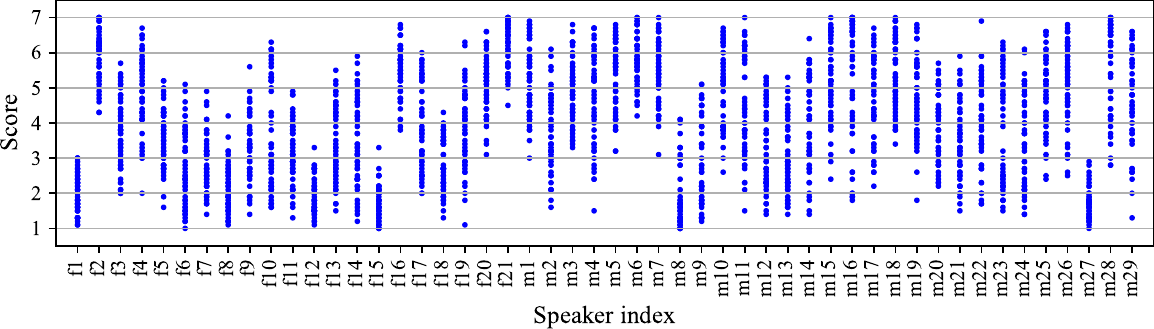}\\
    (a) Separated by speaker\\
    \vspace{10pt}
    \includegraphics[width=1.0\columnwidth]{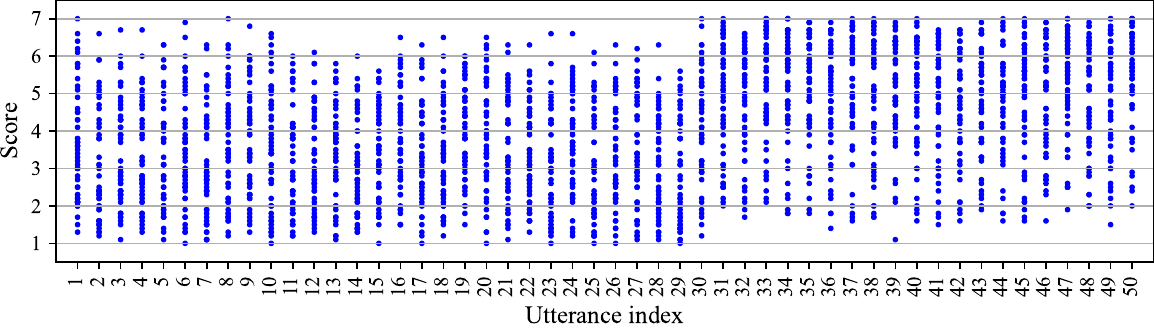}\\
    (b) Separated by utterance
    \caption{Distributions of shout intensity. Each point in this figure represents the average of ten listeners' ratings for a speech sample.}
    \label{fig:subjective_score}
\end{figure}

\subsection{Scoring the shout intensity of each speech sample}
\label{ssec:dataset:score}
We crowdsourced a listening experiment to add shout intensity ratings to the shouted speech samples. 
First, we randomly shuffled the 2,500 shouted speech samples in the dataset and divided them into 125 subsets, each containing 20 speech samples.
The maximum amplitude of each speech sample was normalized to 30,000 to avoid sound-pressure-level-based judgment. 
There was a 200 ms nonspeech interval before and after each speech sample.

To guard against the participation of insincere, unreliable workers who would affect the labeling quality, one dummy speech sample was mixed in the 20 shouted speech samples in each subset. 
This dummy sample was a normal speech utterance of ``hello'' in Japanese by a male speaker. 
Thus, a single crowdsourcing task comprised 21 speech ratings, in which one of the speech samples was a dummy.
Workers who judged the dummy sample to be a shout were considered spammers. 
Each worker was allowed to participate in the experiment no more than three times, and a different subset was assigned for evaluation each time. 
The number of unique workers who participated in this listening experiment was 693.

We asked the workers to wear headphones or earphones and to rate the shout intensity of each speech sample on a seven-point scale from 1 (not a shout at all) to 7 (very shout-like).
The following procedure was applied to collect ten high-quality ratings per speech sample.
We first assigned 12 workers to a single task. 
A worker who scored two or higher for the dummy speech sample in each task was considered a spammer, and all responses from that person were deleted. 
For subsets that received 11 or more ratings, we randomly selected ten of those ratings.
Each speech sample can thus be assigned a single shout intensity score by averaging the ten selected ratings.
In summary, RISC contains 2,500 shouted speech samples that have shout intensity ratings ranging from 1 to 7 in addition to 2,500 normal speech samples.

Figure~\ref{fig:subjective_score} shows scatter plots of the shout intensity ratings obtained in the above listening experiment. 
The dots in the figure represent the average of ten workers' ratings for (a) each speaker and (b) each sentence. 
Specifically, Figure~\ref{fig:subjective_score}~(a) shows the scores of the 50 speech samples uttered by each speaker, where `f' and `m' in the speaker indexes on the horizontal axis represent female and male speakers, respectively. 
On the other hand, Figure~\ref{fig:subjective_score}~(b) shows the scores of the 50 speakers for each sentence. 
The horizontal axis is the sentence index, with 01--05 representing vowel sentences and 06--10, 11--30, and 31--50 representing sentences in the H/L class, L class, and H class, respectively. 
The speaker-specific results in Figure~\ref{fig:subjective_score}~(a) show that speakers f1, m8, and m27 received low scores overall, while speakers f2, f21, and m6 obtained scores higher than the others. 
This indicates that the listeners' perception of the intensity of the shouts varied greatly depending on the speaker. 
In the results by sentence in Figure~\ref{fig:subjective_score}~(b), although the intensities of all sentences tended to vary uniformly, the scores for sentences in the H class tended to be higher than those for the other sentences.
This could be because the linguistic information of these sentences influenced either the listeners or the speakers who uttered the sentences while imagining being in a highly-hazardous situation.

%% file: sec4-proposed.tex
\section{Shout recognition}
\label{sec:proposed}
This paper aims to provide not only a corpus but also comprehensive benchmarks on that corpus.
To this end, this section describes a deep approach for shouted vs. normal speech classification and shout intensity prediction. 
First, we explain the speech features in the spectral and cepstral domains that are used in both conventional methods and the proposed method (see Section~\ref{ssec:feature-extraction}). 
Next, we provide the details of DNN architectures whose inputs are single features (see Section~\ref{ssec:single-network}). 
Finally, we describe our method, in which the outputs of single-feature DNNs are concatenated to yield classification and intensity prediction results (see  Section~\ref{ssec:fusion-network}).

\subsection{Speech feature extraction}
\label{ssec:feature-extraction}
\begin{figure}[ht]
    \centering
    \includegraphics[keepaspectratio, scale=0.44]{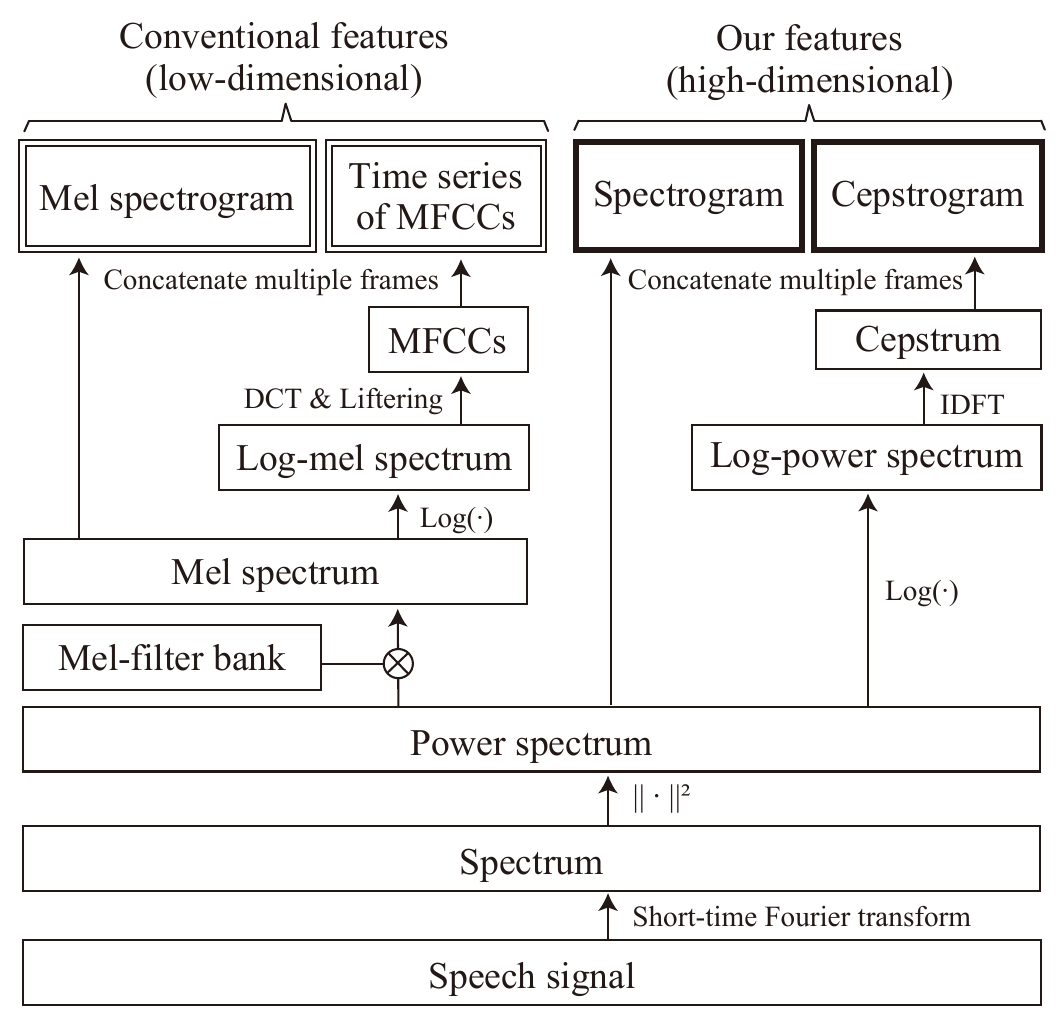}\\
    \caption{Extraction of conventional features and our high-dimensional features.}
    \label{fig:feature-extract}
\end{figure}

For a given audio segment, we partitioned it into successive frames using a Hamming window with a length of 1,024 points (i.e., 64 ms) and a hop length of 512 points (i.e., 32 ms); subsequently, we obtained the features for every 20 frames.
Figure~\ref{fig:feature-extract} summarizes the extraction of these speech features.
It should be noted that neither conventional methods nor our method use the sound pressure level of the input speech as a speech feature for shout recognition, as this feature is highly dependent on the positional relationship between the speaker and the microphone.
Below we provide the details of the MFCCs and the mel spectrogram, which are used in conventional methods~\cite{20-papadimitriou-electronics,20-gaviria-as,20-jasa-baghel,17-mun-icassp,16-laffitte-icassp,09-nanjo-isce,09-ntalampiras-icassp} (hereinafter, \textbf{conventional low-level features}).

\begin{list}{}{}
\item[{\bf Time series of MFCCs (tMFCCs)}:] 
The MFCCs are typical cepstral features. 
Most conventional methods of shouted speech detection used MFCCs with dimensions ranging between 8 and 60~\cite{20-jasa-baghel,20-papadimitriou-electronics,17-mun-icassp,16-laffitte-icassp,11-pohjalainen-icassp,09-ntalampiras-eurasip,09-ntalampiras-icassp,09-nanjo-isce}.
Following~\cite{11-pohjalainen-icassp}, we extracted 30-dimensional MFCCs from each frame and concatenated the vectors over 20 frames, resulting in a 600-dimensional cepstral feature vector.

\item[{\bf Mel spectrogram}:] 
The mel spectrogram belongs to the spectral domain and has been used in recent studies pertaining to sound event detection~\cite{17-eusipco-valenti,16-schroder-dcase}, with dimensions ranging between 25 and 40.
We extracted a 30-dimensional mel spectrogram, whose number of dimensions is the same as that of the MFCCs.\\
\end{list}
Herein, we propose learning features that are suitable for shouted vs. normal speech classification instead of using the conventional features extracted as described above.
The features used in this study (hereinafter, \textbf{high-level features}) are described below.
\begin{list}{}{}
\item[{\bf Spectrogram}:]
A spectrogram represents the temporal variation of a spectrum.
Specifically, applying the short-time Fourier transform to a speech signal yields a 512-dimensional vector of the power spectrum for each frame, and concatenating the vectors of 20 frames results in a 10,240-dimensional spectrogram vector.
Recent studies on sound event detection have used spectrograms as inputs to DNNs and demonstrated their descriptiveness for such target tasks~\cite{21-zhang-applied,20-ciaburro-bdcc,18-ozer-neurocomputing}.
Hence, we used this high-dimensional spectrogram to learn effective spectral features.
\item[{\bf Cepstrogram}:]
Applying the inverse discrete Fourier transform to the log power spectrum yields a cepstrum, and the concatenation of the cepstra of multiple frames yields a cepstrogram.
The cepstrogram represents the temporal variations in the vocal tract and vocal cords.
We set the dimensionality of each cepstrum equal to that of the spectrogram, i.e., 512, resulting in a 10,240-dimensional cepstrogram vector.
\end{list}
The performance of each feature was investigated experimentally.

\subsection{Network architecture}
\label{ssec:single-network}
We used CNN, GRU, and CNN--GRU models to analyze the acoustic and speech features.
We trained these networks as classifiers and regressors using single features.
Figure~\ref{fig:single-networks} shows the architecture of each type of network, whose hyperparameters depend on the number of feature dimensions. 
The detailed settings are as follows:

\begin{figure}[ht]
    \centering
    \includegraphics[keepaspectratio, scale=0.46]{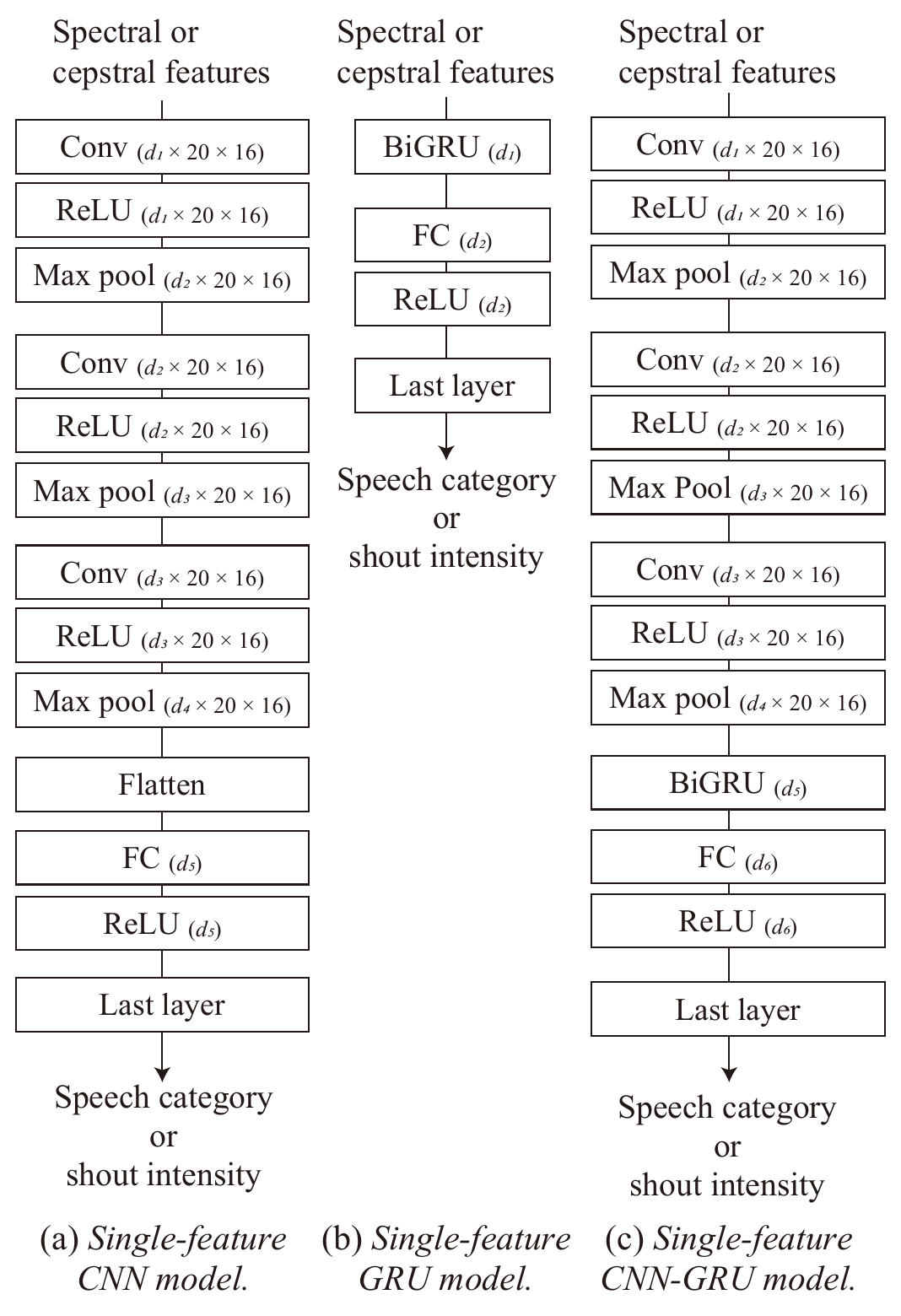}\\
    \caption{Types of single-feature-based networks compared in the experiments. The size of each layer's output is represented in parentheses. The structure of the last layer depends on the target task.}
    \label{fig:single-networks}
\end{figure}

\begin{list}{}{}
\item[{\bf Each single-feature CNN model}] comprised three sets of convolutional and pooling layers followed by two fully connected (FC) layers, as shown in Figure~\ref{fig:single-networks}~(a).
Each of these models treated a set of features collected over 20 frames as an image.
Each convolutional layer contained a $5\times 5$ kernel with a stride of 1, a padding of 2, and 16 channels.
The max pooling layers each contained a $5\times1$ kernel for our high-dimensional features or a $3\times1$ kernel for the conventional low-dimensional features.
The layer parameters $d_1, d_2, d_3, d_4,$ and $d_5$ in the figure were set to $512, 102, 20, 4,$ and $64$ respectively, for high-dimensional input features and $30, 10, 3, 1,$ and $16$ respectively, for low-dimensional input features.

\item[{\bf Each single-feature GRU model}] comprised a bidirectional GRU (BiGRU) layer and two FC layers, as shown in Figure~\ref{fig:single-networks}~(b). 
The input to each of these models was a time series of features from 20 frames.
The layer parameters $d_1$ and $d_2$ in the figure were set to 1,024 and 64, respectively, for high-dimensional input features and 60 and 16, respectively, for low-dimensional input features.

\item[{\bf Each single-feature CNN--GRU model}] comprised three sets of convolutional and pooling layers followed by a BiGRU layer and two FC layers, as shown in Figure~\ref{fig:single-networks}~(c).
Each of these models took feature images as inputs, and the output of the third max pooling layer was passed to the BiGRU layer as a time series of frame features. 
We set the parameters of the convolutional and pooling layers (i.e., $d_1$ to $d_5$ in the figure) to the same values as those in the single-feature CNNs.
The remaining parameter, $d_6$, was set to 64 or 16 for high-dimensional or conventional low-dimensional features, respectively.
\end{list}
Each network used rectified linear units (ReLUs) as activation functions in each layer.
The structure of the last layer in Figure~\ref{fig:single-networks} and the loss function both differed between the speech type classification task and the shout intensity prediction task. 

\subsection{Spectral--cepstral fusion for classification and regression}
\label{ssec:fusion-network}
\begin{figure}[ht]
    \centering
    \includegraphics[keepaspectratio, scale=0.42]{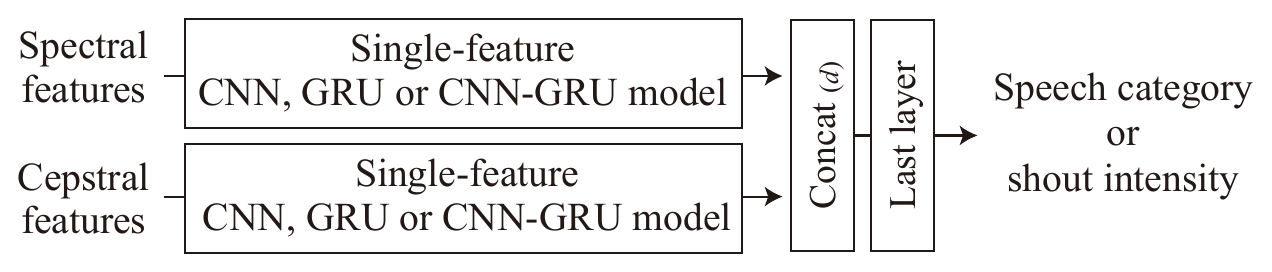}\\
    \caption{Spectral--cepstral fusion network. Note that the output size for each layer is indicated in parentheses. 
    The structure of the last layer depends on the target task.}
    \label{fig:multi-network}
\end{figure}

Our deep spectral--cepstral fusion approach uses features from both domains. 
Figure~\ref{fig:multi-network} shows our DNN architecture, comprising two single-feature networks as described in Section~\ref{ssec:single-network} and an FC layer.
First, we pretrained the single-feature-based networks using either spectral or cepstral features.
Subsequently, we concatenated the outputs from the last ReLU layers of these two single-feature networks and input them into the FC layer.
The number of dimensions of the concatenated features, $d$, was $128$ for high-dimensional features and $32$ for low-dimensional ones.
The concatenated features were then passed to the last layer to obtain the final classification-or-prediction result.
We fine-tuned the entire network using a training dataset, resulting in a feature extractor specific to either shouted vs. normal speech classification or shout intensity prediction.

%% file: sec5-evaluation.tex
\section{Recognition results}
\label{sec:evaluation}
We conducted three experiments using RISC. 
In Experiment~1, the test speech samples were classified into two classes: normal and shouted speech. 
In Experiment~2, the input speech samples were classified into four categories: normal speech (Normal) and three types of shouted speech (Shout-H, Shout-L, and Shout-H/L). 
In Experiment~3, we predicted the shout intensity shown in Figure~\ref{fig:subjective_score}. 
Experiment~1 focused on the general task of conventional shouted speech detection problems, while Experiments~2 and 3 focused on the detection of urgent and critical situations considering the demands of practical surveillance systems.

\subsection{Common settings}

Throughout the experiments, each speech sample in the corpus was downsampled from a sampling frequency of 48 kHz to 16 kHz. 
To consider different noise conditions in the tests, we used NOISE-X92~\cite{93-varga-sc} to add factory noise at the following eight signal-to-noise ratios (SNRs): $\infty$, $20$, $10$, $5$, $0$, $-5$, $-10$, and $-20$ dB.

We implemented the networks shown in Figures~\ref{fig:single-networks} and \ref{fig:multi-network} using PyTorch.
All networks were trained using the Adam optimizer~\cite{14-kingma-arxiv} with an initial learning rate of 0.0001 and momentum parameters of 0.9 and 0.999 on two NVIDIA RTX A6000 GPUs.
The batch size was 256, and 100 epochs were used for training.
Fivefold cross-validation was performed by partitioning the corpus into 40 training-verification speakers and ten test speakers.
In addition to the features described in Section~\ref{sec:proposed}, we also tested MFCCs and their second derivatives, MFCCs\_$\Delta\Delta$, as used in~\cite{20-jasa-baghel} and the original network architecture of the cited work.
As performance measures, we used the F1-score, the weighted F1-score, and the root mean square error (RMSE) for the binary classification, the four-class classification, and the intensity prediction tasks, respectively.

\subsection{Experiment 1: Binary classification}
\label{ssec:evaluation:exp1}
We regarded the 2,500 normal speech samples and the 2,500 shouted speech samples as negative and positive examples, respectively.
The last layer in Figures~\ref{fig:single-networks} and \ref{fig:multi-network} was designed as a combination of an FC layer, FC~(1), and a sigmoid function, Sigmoid~(1); this layer classified the input speech as shouted speech if the output from the sigmoid function was greater than 0.5 and as normal speech otherwise. The mean squared error (MSE) was used as the loss function to train the network.

Table~\ref{tbl:result-exp1} shows the comprehensive evaluation results obtained with the different types of features and network architectures under the eight SNR conditions, and the average F1-scores are provided as well.
The symbol ``$+$'' in the table represents the use of the two corresponding features in a fusion network of the form shown in Figure~\ref{fig:multi-network}.
Among the network architectures, the CNNs achieved higher F1-score than the other architectures.
Focusing on the performance of the single features, we find that the high-level features (i.e., spectrogram or cepstrogram features) achieved better F1-score than the conventional low-level features in the same domain (i.e., the mel spectrogram or MFCCs).
In particular, with a decrease in the SNR, the conventional low-level features suffered a sharp decrease in the F1-score, more so than our high-level features.

\begin{table}[ht]
    \caption{F1-scores obtained with different combinations of features and DNN architectures in Experiment 1.}
    \vspace{5pt}
    \label{tbl:result-exp1}
    \footnotesize
    \centering
        \begin{tabular}{l|l||cccccccc||c}\hline\hline
            & & \multicolumn{8}{c||}{SNR [dB]} & \\\cline{3-10}
            \multicolumn{1}{c|}{Speech features} & \multicolumn{1}{c||}{Model} & $\infty$ & $20$ & $10$ & $5$ & $0$ & $-5$ & $-10$ & $-20$ & Avg.\\\hline
            MFCCs\_$\Delta\Delta$~\cite{20-jasa-baghel} & ~~~DNN & 0.932 & 0.915 & 0.890 & 0.859 & 0.811 & 0.743 & 0.653 & 0.525 & 0.791\\\hline
            Mel spectrogram~\cite{17-eusipco-valenti} & \multirow{6}*{~~~CNN} & 0.960 & 0.960 & 0.959 & 0.958 & 0.954 & 0.941 & 0.873 & 0.573 & 0.897\\ 
            tMFCCs~\cite{17-mun-icassp, 16-laffitte-icassp} &  & 0.963 & 0.962 & 0.958 & 0.955 & 0.942 & 0.925 & 0.856 & 0.638 & 0.900\\
            Mel spectrogram + tMFCCs~\cite{20-gaviria-as} &  & 0.963 & 0.963 & 0.961 & 0.960 & 0.957 & 0.943 & 0.869 & 0.544 & 0.895\\
            \lbrack \textbf{Ours}\rbrack~Spectrogram &   & \textbf{0.968} & 0.967 & 0.967 & 0.966 & 0.964 & 0.962 & 0.929 & 0.665 & 0.924\\
            \lbrack \textbf{Ours}\rbrack~Cepstrogram &   & 0.971 & \textbf{0.970} & \textbf{0.971} & \textbf{0.970} & 0.964 & 0.947 & 0.887 & 0.514 & 0.899\\
            \lbrack \textbf{Ours}\rbrack~Spectrogram + Cepstrogram &  & 0.970 & 0.969 & 0.968 & 0.968 & \textbf{0.966} & \textbf{0.964} & \textbf{0.932} & 0.666 & 0.925\\\hline
            Mel spectrogram~\cite{17-eusipco-valenti} & \multirow{6}*{~~~GRU} & 0.947 & 0.947 & 0.944 & 0.936 & 0.908 & 0.878 & 0.818 & 0.707 & 0.886\\
            tMFCCs~\cite{17-mun-icassp, 16-laffitte-icassp} &  & 0.906 & 0.879 & 0.860 & 0.855 & 0.855 & 0.855 & 0.855 & \textbf{0.855} & 0.865\\
            Mel spectrogram + tMFCCs~\cite{20-gaviria-as} &  & 0.955 & 0.952 & 0.948 & 0.942 & 0.922 & 0.900 & 0.856 & 0.807 & 0.910\\
            \lbrack \textbf{Ours}\rbrack~Spectrogram & & 0.962 & 0.962 & 0.963 & 0.962 & 0.958 & 0.945 & 0.912 & 0.815 & \textbf{0.935}\\
            \lbrack \textbf{Ours}\rbrack~Cepstrogram & & 0.962 & 0.960 & 0.957 & 0.951 & 0.942 & 0.918 & 0.851 & 0.695 & 0.904\\
            \lbrack \textbf{Ours}\rbrack~Spectrogram + Cepstrogram & & 0.963 & 0.962 & 0.962 & 0.962 & 0.959 & 0.948 & 0.915 & 0.804 & 0.934\\\hline
            Mel spectrogram~\cite{17-eusipco-valenti} & \multirow{6}*{CNN--GRU} & 0.954 & 0.955 & 0.956 & 0.957 & 0.953 & 0.941 & 0.880 & 0.575 & 0.896\\
            tMFCCs~\cite{17-mun-icassp, 16-laffitte-icassp} & & 0.956 & 0.956 & 0.953 & 0.950 & 0.939 & 0.923 & 0.859 & 0.685 & 0.903\\
            Mel spectrogram + tMFCCs~\cite{20-gaviria-as} &  & 0.958 & 0.957 & 0.957 & 0.958 & 0.954 & 0.943 & 0.879 & 0.565 & 0.896\\
            \lbrack \textbf{Ours}\rbrack~Spectrogram & & 0.965 & 0.966 & 0.965 & 0.964 & 0.962 & 0.956 & 0.926 & 0.643 & 0.919\\
            \lbrack \textbf{Ours}\rbrack~Cepstrogram & & \textbf{0.968} & 0.969 & 0.968 & 0.967 & 0.964 & 0.950 & 0.895 & 0.574 & 0.907\\
            \lbrack \textbf{Ours}\rbrack~Spectrogram + Cepstrogram &  & 0.966 & 0.966 & 0.966 & 0.966 & 0.963 & 0.957 & 0.924 & 0.602 & 0.914\\\hline\hline
        \end{tabular}
\end{table}

\subsection{Experiment 2: Four-class classification}
\label{ssec:evaluation:exp2}

Next, we experimented with a four-class classifier using the following labels: Normal, Shout-H, Shout-L, and Shout-H/L. 
The last layer of the networks consisted of an FC layer, FC~(4), and a softmax function Softmax~(4). 
The largest output from the softmax function indicated the classification result. 
Cross-entropy was used as the loss function.

\begin{figure}[ht]
    \centering
    \includegraphics[width=0.33\columnwidth]{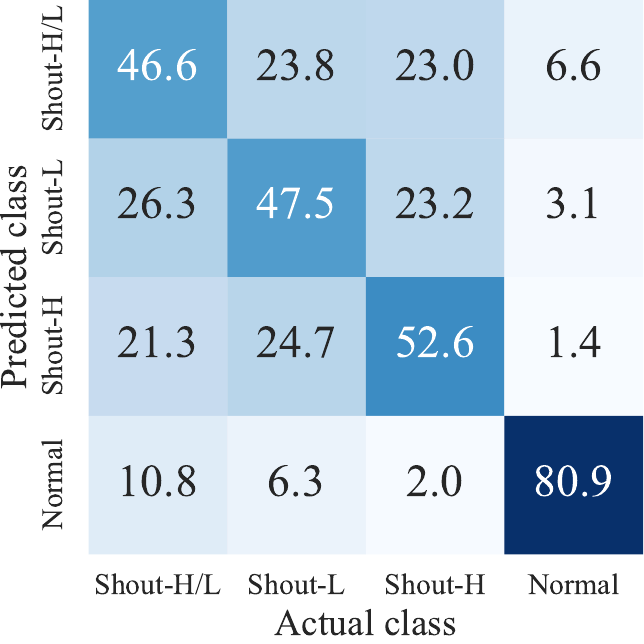}
    \caption{Confusion matrix (model: CNN--GRU, SNR: 0~dB, features: ``Spectrogram + Cepstrogram'').}
    \label{fig:exp2-confusion-matrix}
\end{figure}

Table~\ref{tbl:result-exp2} summarizes the weighted F1-scores of the classification results for normal speech and three types of shouted speech. 
Experiment 2 addressed four-class classification, which is more difficult than the task addressed in Experiment 1, and the weighted F1-score decreased overall.
However, Table~\ref{tbl:result-exp2} shows a clear tendency for our high-dimensional features to show superior performance compared to the low-dimensional features.  
Furthermore, the combination of spectral and cepstral features (i.e., ``Spectrogram + Cepstrogram'') performed the best except for an SNR of -20 dB. 
Regarding the classification network, higher weighted F1-score were achieved at SNRs above 10 dB and below 5 dB by GRU and CNN--GRU models, respectively.
To analyze the classification results in more detail, Figure~\ref{fig:exp2-confusion-matrix} shows the confusion matrix for the CNN--GRU model with an SNR of 0 dB and Spectrogram + Cepstrogram features. 
We can see in this matrix that the normal speech (Normal) could be separated from the shouted speech (Shout-H, Shout-L, and Shout-H/L) with more than 80\% accuracy. 
On the other hand, the levels of discrimination accuracy for shouted speech in the three shout classes, i.e., Shout-H, Shout-L, and Shout-H/L, were 52.6\%, 47.5\%, and 46.6\%, respectively.
These findings reflect the difficulty of classifying shout types based only on acoustic features.
In future work, we should investigate whether the linguistic information that can be obtained through automatic speech recognition can improve the shout type classification performance.

\begin{table}[ht]
    \caption{Weighted F1-scores obtained with different combinations of features and DNN architectures in Experiment 2.}
    \vspace{5pt}
    \label{tbl:result-exp2}
    \footnotesize
    \centering
        \begin{tabular}{l|l||cccccccc||c}\hline\hline
            & & \multicolumn{8}{c||}{SNR [dB]} & \\\cline{3-10}
            \multicolumn{1}{c|}{Speech features} & \multicolumn{1}{c||}{Model} & $\infty$ & $20$ & $10$ & $5$ & $0$ & $-5$ & $-10$ & $-20$ & Avg.\\\hline
            MFCCs\_$\Delta\Delta$~\cite{20-jasa-baghel} & ~~~DNN & \textbf{0.611} & 0.581 & 0.548 & 0.519 & 0.479 & 0.426 & 0.365 & \textbf{0.299} & 0.479\\\hline
            Mel spectrogram~\cite{17-eusipco-valenti} & \multirow{6}*{~~~CNN} & 0.536 & 0.535 & 0.534 & 0.532 & 0.525 & 0.502 & 0.427 & 0.232 & 0.478\\ 
            tMFCCs~\cite{17-mun-icassp, 16-laffitte-icassp} &  & 0.573 & 0.562 & 0.555 & 0.543 & 0.523 & 0.494 & 0.422 & 0.235 & 0.488\\
            Mel spectrogram + tMFCCs~\cite{20-gaviria-as} &  & 0.565 & 0.559 & 0.555 & 0.548 & 0.537 & 0.515 & 0.443 & 0.239 & 0.495\\
            \lbrack \textbf{Ours}\rbrack~Spectrogram &   & 0.539 & 0.539 & 0.537 & 0.534 & 0.527 & 0.516 & 0.483 & 0.278 & 0.494\\
            \lbrack \textbf{Ours}\rbrack~Cepstrogram &   & 0.553 & 0.544 & 0.528 & 0.521 & 0.510 & 0.478 & 0.428 & 0.230 & 0.474\\
            \lbrack \textbf{Ours}\rbrack~Spectrogram + Cepstrogram &  & 0.586 & 0.582 & 0.573 & 0.568 & 0.553 & 0.536 & 0.486 & 0.262 & 0.518\\\hline
            Mel spectrogram~\cite{17-eusipco-valenti} & \multirow{6}*{~~~GRU} & 0.540 & 0.540 & 0.537 & 0.530 & 0.495 & 0.443 & 0.351 & 0.187 & 0.453\\
            tMFCCs~\cite{17-mun-icassp, 16-laffitte-icassp} &  & 0.468 & 0.397 & 0.295 & 0.238 & 0.198 & 0.172 & 0.152 & 0.150 & 0.259\\
            Mel spectrogram + tMFCCs~\cite{20-gaviria-as} &  & 0.540 & 0.538 & 0.536 & 0.527 & 0.498 & 0.465 & 0.399 & 0.247 & 0.468\\
            \lbrack \textbf{Ours}\rbrack~Spectrogram & & 0.602 & 0.599 & 0.593 & 0.586 & 0.570 & 0.532 & 0.440 & 0.276 & 0.525\\
            \lbrack \textbf{Ours}\rbrack~Cepstrogram & & 0.587 & 0.564 & 0.530 & 0.518 & 0.470 & 0.399 & 0.306 & 0.161 & 0.442\\
            \lbrack \textbf{Ours}\rbrack~Spectrogram + Cepstrogram & & \textbf{0.611} & \textbf{0.606} & \textbf{0.597} & 0.591 & 0.578 & 0.541 & 0.457 & 0.281 & 0.533\\\hline
            Mel spectrogram~\cite{17-eusipco-valenti} & \multirow{6}*{CNN--GRU} & 0.544 & 0.544 & 0.544 & 0.541 & 0.537 & 0.498 & 0.421 & 0.232 & 0.483\\
            tMFCCs~\cite{17-mun-icassp, 16-laffitte-icassp} & & 0.577 & 0.563 & 0.547 & 0.527 & 0.511 & 0.478 & 0.422 & 0.244 & 0.484\\
            Mel spectrogram + tMFCCs~\cite{20-gaviria-as} &  & 0.560 & 0.560 & 0.556 & 0.552 & 0.545 & 0.509 & 0.439 & 0.255 & 0.497\\
            \lbrack \textbf{Ours}\rbrack~Spectrogram & & 0.589 & 0.590 & 0.587 & 0.584 & 0.574 & 0.540 & 0.480 & 0.267 & 0.526\\
            \lbrack \textbf{Ours}\rbrack~Cepstrogram & & 0.536 & 0.542 & 0.536 & 0.530 & 0.521 & 0.502 & 0.433 & 0.237 & 0.480\\
            \lbrack \textbf{Ours}\rbrack~Spectrogram + Cepstrogram &  & 0.597 & 0.599 & 0.596 & \textbf{0.592} & \textbf{0.582} & \textbf{0.547} & \textbf{0.491} & 0.269 & \textbf{0.534}\\\hline\hline
        \end{tabular}
\end{table}

\subsection{Experiment 3: Shout intensity prediction}
\label{ssec:evaluation:exp3}

Finally, we trained a regressor using the 2,500 shouted speech samples collected for all 50 sentences and their shout intensities.
The last layer of each network consisted of only an FC layer, FC~(1). 
Since the ratings ranged from 1 to 7, we bounded the outputs of FC~(1) to this range.
We used the MSE as the loss function.
 
\begin{figure}[ht]
    \centering
    \includegraphics[width=0.33\columnwidth]{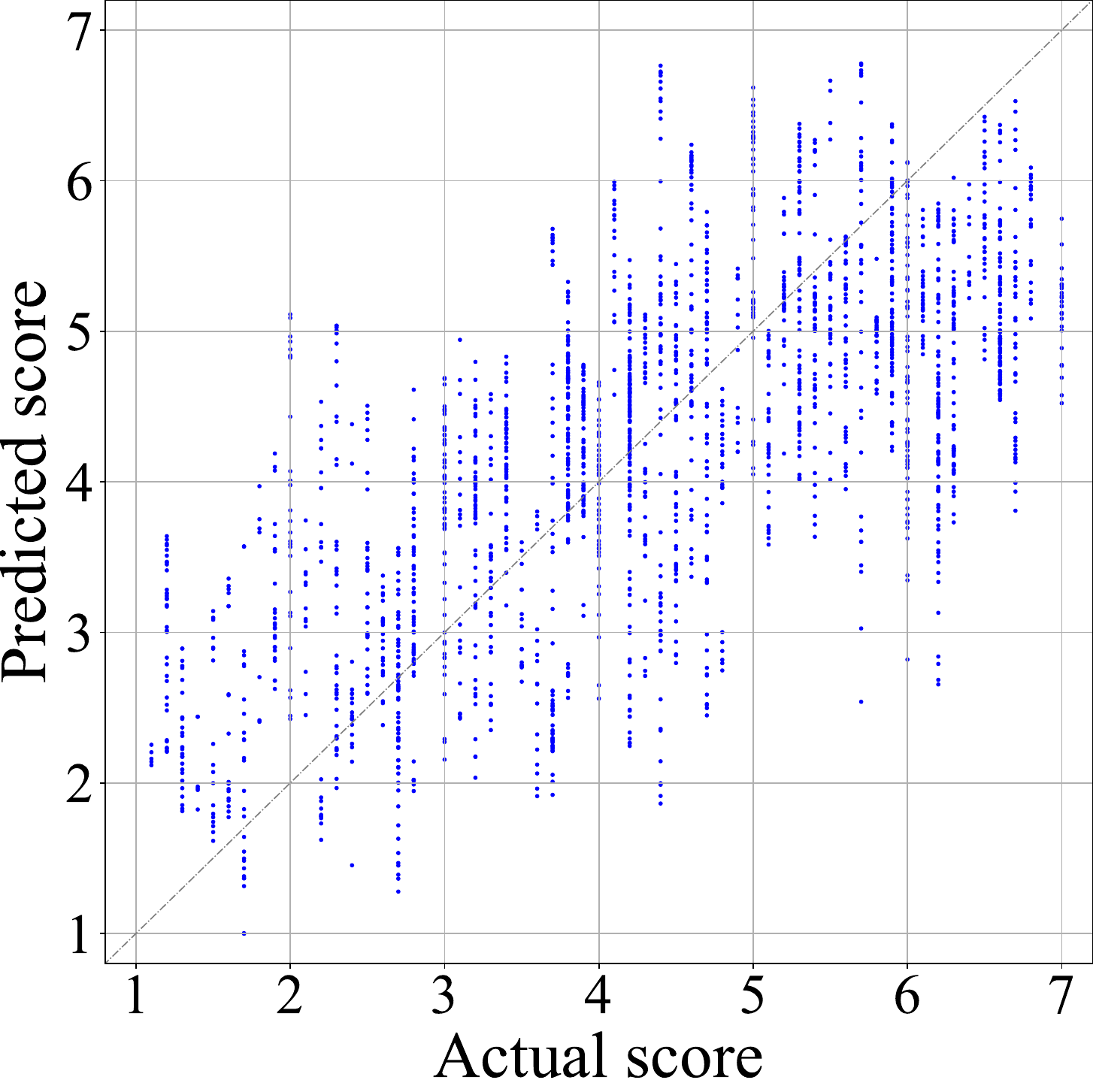}
    \caption{Scatter plot (model: CNN, SNR: 0~dB, features: ``Spectrogram + Cepstrogram'').}
    \label{fig:exp3-scatter-plot}
\end{figure}

Table~\ref{tbl:result-exp3} reports the RMSEs between the actual and predicted shout intensity values. 
We can see from the results that the prediction error with the high-dimensional features was reduced compared with that obtained with the conventional low-dimensional features.
Among all network architectures, the CNNs achieved the lowest RMSEs under most SNR conditions. 
The CNN model with Spectrogram + Cepstrogram features achieved the best performance averaged over all SNRs.
Figure~\ref{fig:exp3-scatter-plot} presents a scatter plot produced for the CNN model with Spectrogram + Cepstrogram features that shows the relationship between the actual and predicted values. 
Although a positive correlation is evident, there is still room for improvement in the prediction accuracy even when we apply feature learning based on the spectral and cepstral domains.
These results demonstrate that our new corpus RISC presents a challenging task for research on shout detection.

\begin{table}[ht]
    \caption{RMSEs obtained with different combinations of features and DNN architectures in Experiment 3.}
    \vspace{5pt}
    \label{tbl:result-exp3}
    \footnotesize
    \centering
        \begin{tabular}{l|l||cccccccc||c}\hline\hline
            & & \multicolumn{8}{c||}{SNR [dB]} & \\\cline{3-10}
            \multicolumn{1}{c|}{Speech features} & \multicolumn{1}{c||}{Model} & $\infty$ & $20$ & $10$ & $5$ & $0$ & $-5$ & $-10$ & $-20$ & Avg.\\\hline
            MFCCs\_$\Delta\Delta$~\cite{20-jasa-baghel} & ~~~DNN & 1.471 & 1.541 & 1.594 & 1.630 & 1.670 & 1.706 & 1.731 & 1.752 & 1.637\\\hline
            Mel spectrogram~\cite{17-eusipco-valenti} & \multirow{6}*{~~~CNN} & 1.372 & 1.373 & 1.375 & 1.370 & 1.399 & 1.433 & 1.513 & 1.725 & 1.445\\ 
            tMFCCs~\cite{17-mun-icassp, 16-laffitte-icassp} &  & \textbf{0.985} & 1.080 & 1.130 & 1.185 & 1.243 & 1.374 & 1.476 & 1.673 & 1.268\\
            Mel spectrogram + tMFCCs~\cite{20-gaviria-as} &  & 1.160 & 1.201 & 1.239 & 1.265 & 1.320 & 1.398 & 1.495 & 1.735 & 1.352\\
            \lbrack \textbf{Ours}\rbrack~Spectrogram &   & 1.110 & 1.108 & 1.117 & 1.141 & 1.202 & 1.313 & 1.451 & 1.686 & 1.266\\
            \lbrack \textbf{Ours}\rbrack~Cepstrogram &   & 1.039 & 1.072 & 1.089 & 1.130 & 1.173 & 1.307 & 1.437 & \textbf{1.662} & 1.239\\
            \lbrack \textbf{Ours}\rbrack~Spectrogram + Cepstrogram &  & 1.027 & \textbf{1.037} & \textbf{1.056} & \textbf{1.089} & \textbf{1.145} & \textbf{1.260} & \textbf{1.405} & 1.691 & \textbf{1.214}\\\hline
            Mel spectrogram~\cite{17-eusipco-valenti} & \multirow{6}*{~~~GRU} & 1.457 & 1.457 & 1.462 & 1.467 & 1.479 & 1.512 & 1.607 & 1.715 & 1.520\\
            tMFCCs~\cite{17-mun-icassp, 16-laffitte-icassp} &  & 2.125 & 2.032 & 2.055 & 2.072 & 2.078 & 2.091 & 2.085 & 2.077 & 2.077\\
            Mel spectrogram + tMFCCs~\cite{20-gaviria-as} &  & 1.386 & 1.425 & 1.448 & 1.448 & 1.443 & 1.498 & 1.594 & 1.656 & 1.487\\
            \lbrack \textbf{Ours}\rbrack~Spectrogram & & 1.188 & 1.193 & 1.213 & 1.250 & 1.289 & 1.414 & 1.580 & 1.741 & 1.358\\
            \lbrack \textbf{Ours}\rbrack~Cepstrogram & & 1.196 & 1.189 & 1.265 & 1.340 & 1.421 & 1.520 & 1.560 & 1.657 & 1.394\\
            \lbrack \textbf{Ours}\rbrack~Spectrogram + Cepstrogram & & 1.145 & 1.165 & 1.196 & 1.236 & 1.277 & 1.391 & 1.543 & 1.723 & 1.334\\\hline
            Mel spectrogram~\cite{17-eusipco-valenti} & \multirow{6}*{CNN--GRU} & 1.364 & 1.365 & 1.372 & 1.394 & 1.422 & 1.430 & 1.495 & 1.673 & 1.439\\
            tMFCCs~\cite{17-mun-icassp, 16-laffitte-icassp} & & 0.994 & 1.059 & 1.116 & 1.173 & 1.248 & 1.368 & 1.481 & 1.678 & 1.265\\
            Mel spectrogram + tMFCCs~\cite{20-gaviria-as} &  & 1.191 & 1.220 & 1.250 & 1.287 & 1.347 & 1.404 & 1.500 & 1.726 & 1.366\\
            \lbrack \textbf{Ours}\rbrack~Spectrogram & & 1.256 & 1.261 & 1.277 & 1.294 & 1.333 & 1.422 & 1.544 & 1.798 & 1.398\\
            \lbrack \textbf{Ours}\rbrack~Cepstrogram & & 1.029 & 1.064 & 1.099 & 1.148 & 1.212 & 1.368 & 1.492 & 1.712 & 1.265\\
            \lbrack \textbf{Ours}\rbrack~Spectrogram + Cepstrogram &  & 1.195 & 1.207 & 1.224 & 1.242 & 1.272 & 1.368 & 1.488 & 1.826 & 1.353\\\hline\hline
        \end{tabular}
\end{table}

%% file: sec6-conclusion.tex
\section{Conclusion}
\label{sec:conclusion}
This paper has presented RISC, a new corpus comprising diverse shouted speech samples, such as angry shouts, screams, and cheers, along with their shout type and intensity information. 
We have described a detailed pipeline for corpus construction, which has mostly not been specified to date in the literature on shouted speech detection.
To provide a comprehensive performance comparison between deep approaches as a benchmark, we performed experiments focusing on two speech type classification tasks and an intensity prediction task.
From the results achieved using various combinations of network architectures and speech features, we observed that feature learning based on spectrograms and cepstrograms achieved high performance on all three tasks, no matter which network architecture is used.

We also found that shout type classification and intensity prediction, which have not been addressed in previous studies, are still challenging even for the high-dimensional feature learning approach.
In future work, we should improve the performance on these tasks by developing effective deep architectures.
Another possible strategy is to introduce linguistic information obtained through automatic speech recognition.
Thus, toward the construction of sophisticated audio surveillance systems, research on shouted speech detection needs to be integrated with natural language processing.